\documentclass[aps,prd,twocolumn,nofootinbib,floatfix,unsortedaddress]{revtex4-1}
\DeclareSymbolFont{matha}{OML}{txmi}{m}{it}
\DeclareMathSymbol{\varv}{\mathord}{matha}{118 }
\usepackage{setspace}
\usepackage{graphicx}
\usepackage{dcolumn}
\usepackage{multirow}
\usepackage{stackrel}
\usepackage{subfigure}
\usepackage{times,mathptm}
\usepackage{float}
\usepackage{color}
\usepackage{amsmath,amsfonts}
\usepackage{mathptmx}
\usepackage{mathrsfs}
\usepackage{bbm}
\usepackage{bm}
\usepackage{xfrac}

\newcommand{\vv}{\upsilon}

\newcommand{\beq}{\begin{equation}}
\newcommand{\eeq}{\end{equation}} 
\newcommand{\bea}{\begin{eqnarray}}
\newcommand{\eea}{\end{eqnarray}}

\def\lsim{\mathrel{\rlap{\lower4pt\hbox{\hskip1pt$\sim$}}
    \raise1pt\hbox{$<$}}}
\def\gsim{\mathrel{\rlap{\lower4pt\hbox{\hskip1pt$\sim$}}
    \raise1pt\hbox{$>$}}}
\newcommand{\adot}{\dot{a}}
\newcommand{\pdot}{\partial_t \vph}
\newcommand{\vp}{\vec{\vph}}
\newcommand{\vB}{\vec{B}}
\newcommand{\vabla}{\vec{\nabla}}

\renewcommand{\d}{\delta}

\renewcommand{\l}{\lambda}

\newcommand{\p}{\phi}
\renewcommand{\b}{\beta}
\renewcommand{\a}{\alpha}

\newcommand{\n}{\nu}
\newcommand{\m}{\mu}

\newcommand{\g}{\gamma}
\renewcommand{\r}{\rho}
\newcommand{\e}{\epsilon}
\newcommand{\s}{\sigma}

\renewcommand{\th}{\theta}

\newcommand{\vph}{\varphi}
\newcommand{\oh}{\frac{1}{2}}
\newcommand{\oq}{\frac{1}{4}}

\newcommand{\non}{\nonumber}

\newcommand{\rf}[1]{(\ref{#1})}

\newcommand{\pa}{\partial}
\renewcommand{\vec}[1]{\bm #1}

\usepackage{ulem}

\begin{document}

\title{Inflation in an external four-form gauge field} 


\bigskip
\bigskip

\author{Jeff Greensite}

\bigskip

\affiliation{ \vspace{5pt} Physics and Astronomy Department, San Francisco State
University,   \\ San Francisco, CA~94132, USA}
 
\date{\today}
\vspace{60pt}
\begin{abstract}

\singlespacing

     We consider the possibility that the universe, viewed as a three-brane, originated in a region of strong external field strength due to a
four-form gauge field in the bulk. It is shown that in a scenario of this kind inflation is generic for a wide range of initial conditions.  This is true even for small field inflation with a simple quadratic inflaton potential, as well as for Higgs potentials with the initial field well away from the local maximum, not necessarily starting from rest.  The power spectrum, spectral index, and $r$ parameter help to constrain parameters in this scenario, and $r<0.1$ favors Higgs potentials.

\end{abstract}

%
%
%
\maketitle

\singlespacing
\section{\label{intro}Introduction}

    The view that our universe should be viewed as a 3-brane in higher dimensions has been advocated, in various forms, by
many authors, e.g.\ \cite{Regge,ArkaniHamed:1998nn,Randall:1999vf,Alishahiha:2004eh,HenryTye:2006uv,Dvali:2000hr,
Pavsic:2000ws,*Pavsic:2000qy}.   But three branes couple in a natural way to an abelian four-form gauge field in the bulk.  If the embedding coordinates of the brane are denoted $\phi^a(x)$, and the brane is charged with respect to the gauge field, then the corresponding interaction term is
\beq
S_A = {q_0\over 4!} \int d^4x ~ A_{abcd}[\p(x)] \e^{\a\b\g\d} \pa_\a \p^a \pa_\b \p^b \pa_\g \p^c \pa_\d \p^d \ .
\eeq 
It is interesting to ask what effect a strong external 4-form gauge field might have had on the dynamics of the early universe, if the universe were exposed to such a field at that early time.

   Of course such a question can only be answered in the context of a specific model.  The model I will consider here was proposed
in \cite{Greensite:2016dhu}.  It consists of $S_A$ plus the usual action of the standard model fields.  In addition there is the Einstein-Hilbert action, where the metric depends on the embedding, and an inflaton action.  What is a little non-standard is that the inflaton fields
are regarded as a subset of the $D+1+N$ embedding coordinates, while the induced metric $g_{\m \n}$ is taken to depend only on the remaining
embedding coordinates.  Explicitly
\bea
    S &=& S_{SM} + S_A + {1\over 16\pi G}\int d^4x \sqrt{-g} R  
\non \\
& & -  \s^4 \int d^4x \sqrt{-g} \Bigl( \oh g^{\m\n} \pa_\m \p^s \pa_\n \p^s + U(\phi) \Bigr) \ ,
\label{S0}
\eea
where $S_{SM}$ is the action of standard model (and possibly beyond-standard-model) fields, and
\bea
          g_{\m\n} &=& \pa_\m \p^A \eta_{AB} \pa_\n \p^B  ~~~,~~~ A,B = 0,1,...,D
\label{induce}
\eea
is the induced metric of a three brane in a $D+1$-dimensional Minkowski space.   The constant $\s$ has the dimensions of
mass. The remaining $N$ coordinates 
\beq
\{\phi^s, ~ s=D+1,...,D+1+N\}
\eeq
are identified with inflaton fields.  We adopt the convention that upper case Latin indices run from 0 to $D$,
indices $r,s$ run from $D+1$ to $D+N$, and all other lower case Latin indices run from $0$ to $D+N$.  It is
convenient to define
\beq
          \vph^s = \s^2 \phi^s  ~~~\mbox{and}~~~ V[\vph] =  \s^4 U(\phi) \ ,
\eeq
so that the inflaton field and the potential have the conventional dimensions.  We will consider in detail two specific examples, namely a simple quadratic potential
\beq
          V(\vph) = \oh m^2 \vph^s \vph^s \ ,
\label{quadratic}
\eeq
and a Higgs potential 
\beq
        V(\vph) = \l (\vph^s \vph^s - m^2)^2 \ ,
\label{higgs}
\eeq
where, in the latter case, the inflaton starts out at $\vph^s \vph^s < m^2$.

   In section \ref{dynamics} below we review, following ref.\ \cite{Greensite:2016dhu}, the equations of motion of this system, and then specialize to the simplest non-trivial case of a two-component inflaton and a constant external field strength, leading to early-universe dynamical equations for the inflaton zero mode and the metric scale factor.  In section III these equations are simplified to something analogous to slow-roll equations, although in contrast to a slow roll down a potential hill the dynamics results in a kind of spiral motion in inflaton field space towards the minimum of the inflaton potential.  What is going on is that while the inflaton potential tends to pull the inflaton field towards the minimum of the potential, the external field provides a counterbalancing velocity-dependent force analogous to a 
 $\vec{\vv} \times \vec{B}$ force, orthogonal to gravitational friction, away from the minimum.  It is shown that inflation in this model does not require fine tuning of either couplings or initial conditions, even in the small field case.  Scalar field perturbations are considered in section IV,  and the power spectrum, spectral index $n_s$, and $r$-parameter are expressed in terms of the parameters of the model and the initial state.  It has been inferred from the Planck data that the tensor to scalar ratio is $r < 0.1$.  This fact favors the Higgs potential over the quadratic potential in the external field scenario, although both are inflationary at small fields.
   
\section{\label{dynamics} Equations of motion}

    Variation of the action \rf{S0} with respect to the embedding coordinates $\p^A$ leads to equations of motion
\bea
& &2 \eta_{AB} \pa_\m (E^{\m \n} \pa_\n \p^B)   \non \\
& & \qquad - {q_0\over 4!} F_{Aabcd} \e^{\a\b\g\d} \pa_\a \p^a \pa_\b \p^b \pa_\g \p^c \pa_\d \p^d = 0 \; ,
\label{eom1}
\eea
where
\bea
        E^{\m \n} &\equiv& \oh \sqrt{-g} \left\{ -{1\over 8\pi G } G^{\m \n}  + T^{\m \n}   \right\} \; ,
\eea
with $G^{\m \n}, T^{\m \n}$ the Einstein and stress-energy tensors of ${S-S_A}$.  Variation of the action with respect to the inflaton fields $\vph^s$ leads to the equations of motion
\bea
& & \pa_\m(\sqrt{-g} g^{\m\n} \pa_\n \vph^s) - \sqrt{-g} {\pa V \over \pa \vph^s} \non \\
& & \; \; + {q_0 \over 4! \s^2} F_{sabcd}\e^{\a\b\g\d} \pa_\a \p^a \pa_\b \p^b \pa_\g \p^c \pa_\d \p^d = 0 \; ,
\label{eom2}
\eea
where $F_{fabcd}$ is the field strength
\bea
 F_{fabcd}  
  = {\pa A_{abcd} \over \pa \p^f} -  {\pa A_{fbcd} \over \pa \p^a}  +  {\pa A_{facd} \over \pa \p^b} 
-  {\pa A_{fabd} \over \pa \p^c}  +  {\pa A_{fabc} \over \pa \p^d}   \non \\
\label{F}
\eea
corresponding to the four-form gauge field. These equations are of course supplemented by the usual equations of motion of the
standard model fields, which we ignore for now.

    For the purposes of simplified cosmology it is sufficient to assume that the induced metric on the three-brane has the usual
Friedman-Lemaitre form with scale factor $a(t)$, which requires a minimum of five coordinates $\phi^A$ in the bulk.  For zero spatial curvaure, we can choose the embedding \cite{Rosen, LachiezeRey:2000my}
\bea
         \p^0 &=& \oh \left\{ a(t) + \int^t {dt' \over da/dt'} + a(t) r^2 \right\} \non \\
         \p^1 &=& a(t) r \cos(\th)   \non \\
         \p^2 &=& a(t) r \sin(\th) \cos(\chi)   \non \\
         \p^3 &=& a(t) r \sin(\th) \sin(\chi)   \non \\
         \p^4 &=& \oh \left\{ a(t) - \int^t {dt' \over da/dt'} - a(t) r^2 \right\} \; ,
\label{embed}
\eea
and the remaining inflaton coordinates are numbered $\phi^5, \phi^6$.
    
    Next we suppose that there is a constant field strength in the bulk oriented orthogonal to the inflaton plane, with the gauge field  taken to have components
\bea
        A_{5123}[\p] &=& - \oh B \p^6 \; , \non \\
        A_{6123}[\p] &=& \oh B  \p^5 \; .
\label{B}
\eea
The four-form gauge field $A_{abcd}$ is antisymmetric under permutations of indices, but apart from \rf{B} and components obtained from \rf{B} by permutation, it is assumed that all other components vanish.  

     For an initial discussion of inflation in this scenario, we make the usual simplifying assumptions of spatial homogeneity and isotropy, taking in particular
\beq
           \p^{5,6}(x,y,z,t) = \p^{5,6}(t)    \; ,
\label{isotropic}
\eeq
and $\p^a = 0$ for $a>6$.  In conjunction with \rf{B}, this has the consequence that
\beq
        F_{Aabcd} \e^{\a\b\g\d} \pa_\a \p^a \pa_\b \p^b \pa_\g \p^c \pa_\d \p^d = 0 \; .
\eeq
This is because two of the indices $abcd$ must be 5 and 6, so the expression necessarily includes at least
one space derivative of $\vph^s$, which vanishes according to \rf{isotropic}.
Then the equation of motion $\rf{eom1}$ is satisfied by $E^{\m\n}=0$, which are the standard Einstein field
equations.   For a Friedman-Lemaitre metric, disregarding the other standard model fields, the Einstein equations are 
just the conventional expressions for the $a(t)$ scale factor coupled to a pair of scalar fields: 
\bea
            {\adot^2 \over a^2} &=&  {8\pi G \over 3} \left( \oh \pdot^s \pdot^s + V(\vph) \right)  \; , \non \\
            {\ddot{a} \over a}  &=&   {8\pi G \over 3} \left( - \pdot^s \pdot^s + V(\vph) \right) \; .
\label{Einstein}
\eea
The equations of motion for the $\vph^s$, however, involve the field strength
\bea
         \pa^2_t \vph^5 -  qB \pdot^6 + 3 {\adot \over a}\pdot^5 + {\pa V \over \pa \vph^5} &=& 0 \; , \non \\ 
         \pa^2_t \vph^6 +  qB \pdot^5 + 3 {\adot \over a}\pdot^6 + {\pa V \over \pa \vph^6} &=& 0 \; ,
\label{eom3}
\eea
where $q = q_0/\s^4$. It is not hard to verify the consistency of \rf{Einstein} and \rf{eom3}.

\section{\label{spiral}The slow spiral}
  
  Numerical solutions of (\ref{Einstein},~\ref{eom3}), a sample of which are presented in the next section, show that for a very wide range of initial conditions $\vph^s(0), \pdot^s(0)$, field strength $qB$, and parameters $m,\lambda$ in the potential, the evolution of $\vph^s(t)$ rapidly settles into a spiral in the $\vph^5-\vph^6$ plane, drifting slowly towards the origin in the case of the quadratic potential, and towards 
$\vph^2=m^2$ in the case of the Higgs potential. It is the drift towards the minimum, rather than the magnitude of $\pa_t \vph$ itself, which is ``slow'' in this scenario.  

\begin{figure}[t!]
\subfigure[~quadratic potential]  
{   
 \label{quadfree}
 \includegraphics[scale=0.3]{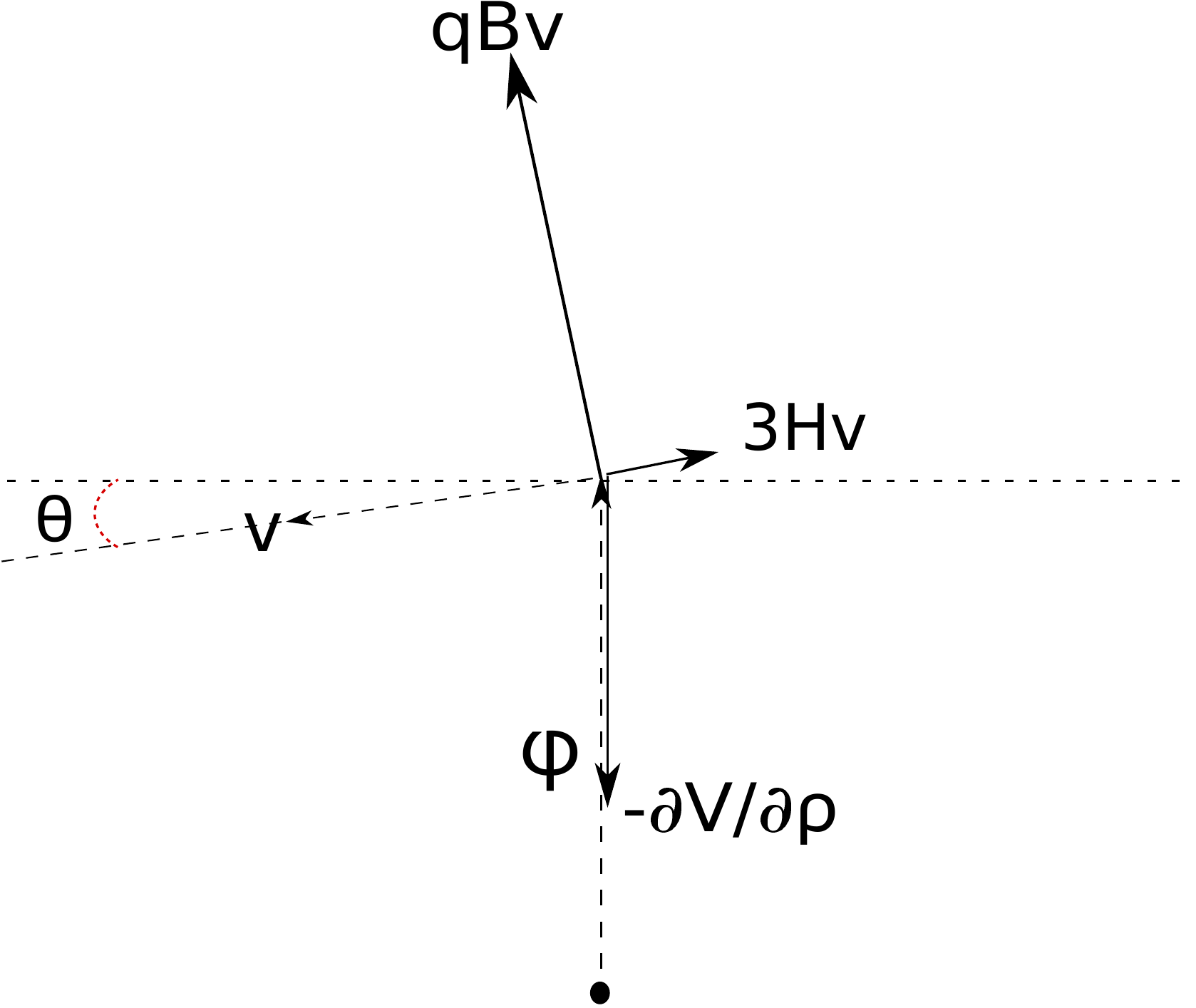}
}
\subfigure[~Higgs potential]  
{   
 \label{higgsfree}
 \includegraphics[scale=0.3]{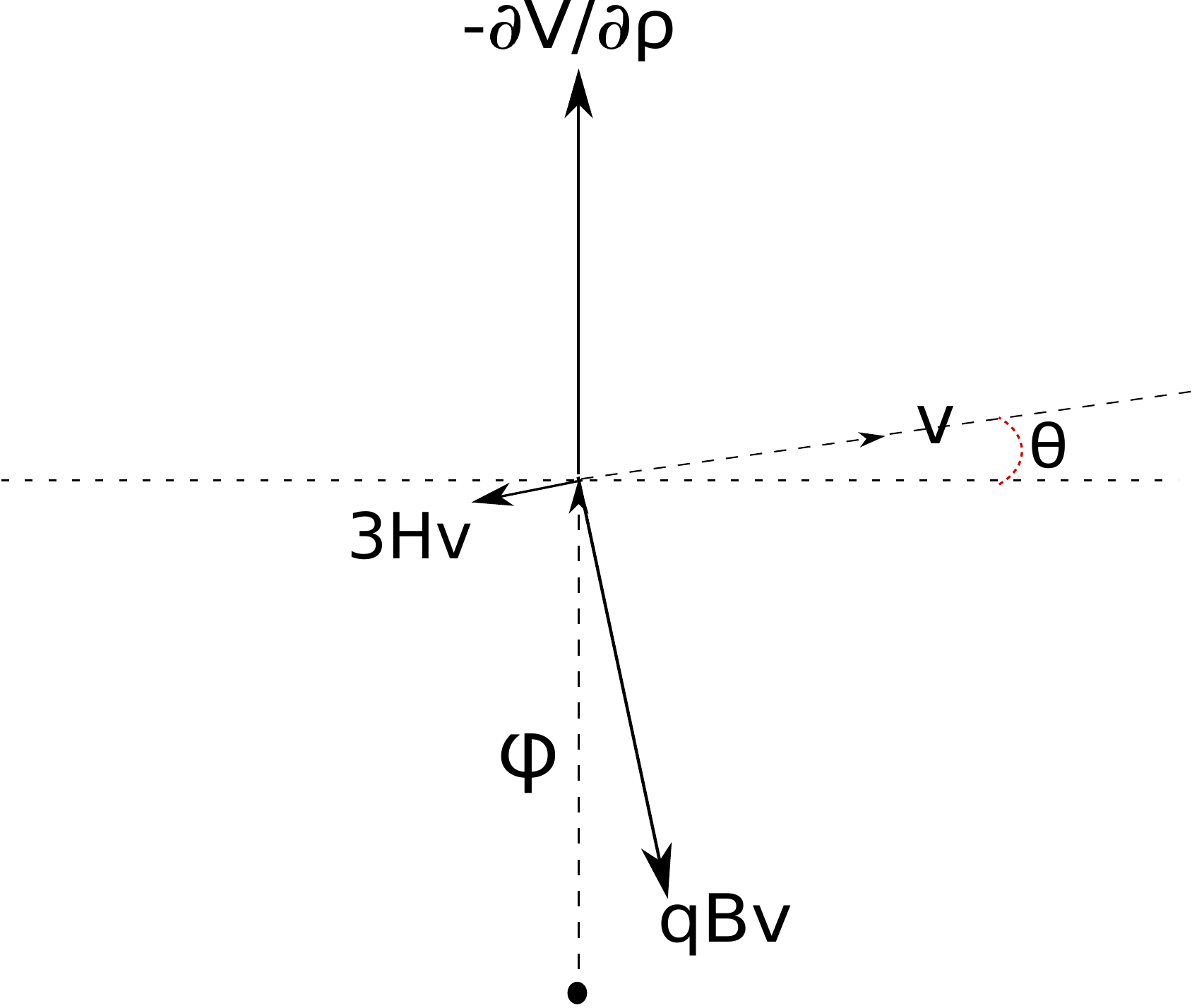}
}
\caption{``Force'' diagrams for the slow spiral approximation, where the spiral is counterclockwise inward towards the origin for the quadratic potential (a), and clockwise outward towards $\r=m$ for the Higgs potential (b).  The approximation is that $\pa_t^2 \vp$ towards the origin (solid dot) is simply the centripetal acceleration required for circular motion in the $\vp$-plane, neglecting any acceleration beyond that.}
\label{freebody}
\end{figure}

    The spiral motion in the $\vph^{5,6}$ plane, with a gradual drift to the minimum, is best understood as a balance of forces in a plane.  Define
\beq
          \vp \equiv \left[\begin{array}{c} 
                           \vph^5 \cr
                           \vph^6 \cr
                           0 \end{array} \right] ~~~,~~~ 
          \vB \equiv \left[\begin{array}{c} 
                           0 \cr
                           0 \cr
                           B \end{array} \right] ~~~,~~~
          \vabla \equiv \left[\begin{array}{c} 
                           {\pa/\pa \vph^5} \cr
                           {\pa/\pa \vph^6} \cr
                           0 \end{array} \right] \ ,
\eeq
and $H=\adot/a$ as usual.  Then the inflaton equations of motion can be written as
\beq
         \pa^2_t \vp +  q\vB \times \pa_t \vp + 3 H\pa_t \vp + \vabla V = 0 \ ,
\eeq         
which can be thought of as the equations of motion of a particle in a plane.  There are three forces on this ``particle'':  a force $-\vabla V$ towards the minimum of the potential, a gravitational drag force in a direction opposite to the
velocity $\pa_t \vp$, and a ``Lorentz force'' in the plane which is perpendicular to the velocity.   In the standard slow roll approximation with a single inflaton field, the second time derivative of the inflaton field is neglected.  In our case the slow roll is in the radial direction, so
the approximation is that the second time derivative can be equated to the centripetal acceleration required for circular motion.
Referring to Fig.\ \ref{freebody} and defining
\bea
            \r &\equiv& |\vp| ~~~,~~~ \vv \equiv |\pa_t \vp|   ~~~,~~~ 
            V'(\r) = {\pa V \over \pa \r} \non \\
            \b &=& \left\{ \begin{array}{rl} 
                       1 & \mbox{Higgs potential} \cr
                      -1 & \mbox{quadratic potential} \end{array} \right. \ , 
\eea
the equations of motion in this ``slow spiral'' approximation are
\bea
         V' + \b (qB \cos\th + 3H \sin\th)\vv&=& {\vv^2\cos^2\th  \over \r} \label{spiral1} \\
          qB \vv\sin\th - 3H \vv \cos\th &=& 0 \label{spiral2} \ .
\eea
Solving \rf{spiral1} for $\vv$, we have
\bea
\vv &=& {\r \b \over 2\cos^2\th}\bigg[(qB\cos\th + 3H\sin\th)  \non \\
& & - \sqrt{(qB\cos\th + 3H\sin\th)^2 + 4V'\cos^2\th/\r} \bigg] \ ,
\label{vv}
\eea
and from \rf{spiral2}
\beq
         \tan\th = {3H \over qB} \ .
\label{slow0}
\eeq
For the cases that we will be concerned with,
\beq
       \th \ll 1 ~~,~~ qB \gg H ~~,~~ (qB)^2 \gg V'/\r \ ,
\eeq 
so that \rf{vv} simplifies to
\beq
          \vv \approx -\b  {V' \over qB} \ .
\label{slow1}
\eeq
In polar coordinates, where
\beq
\vph^5=\r \cos\a ~~~,~~~ \vph^6=\r\sin\a \ ,
\eeq 
the equations of motion are
\bea
          {d\r \over dt} &=& - \vv \sin\th  \approx - 3H {V'\over (qB)^2}  \label{slow2} \\
           {d\a \over dt} &=& \b {\vv \cos\th \over \r}  \approx - {V' \over qB \r}  \label{slow3} \ .
\eea
Assuming that the fractional variation of $H$ is negligible in a period  $\vv/(2\pi \r)$, the solution of these equations is a 
spiral in the $\vph$-plane towards the minimum of the potential, i.e.\ inwards towards $\r=0$ for the quadratic potential or,
if $\r < m$ initially, outwards to $\r=m$ for the Higgs potential.

\section{Numerical Solutions}

   Next we compare the slow spiral equations to numerical solutions of (\ref{Einstein},~\ref{eom3}), which make no assumptions about force balance or the time-dependence of $H$.  In a conventional treatment with $qB=0$, slow roll inflation in a quadratic potential requires a strong initial field $\r_0$ above the Planck scale, $\r_0 > M_P$,  while small field inflation in a Higgs potential calls for an initial $\r_0$ fine tuned to be extremely close to the local maximum, with $\vv_0$ very close to zero.   However, only taking $qB \gg m$, the equations of motion in the external field scenario predict inflationary behavior in either potential without these special conditions.

\begin{figure}[h!]   
 \centerline{\includegraphics[scale=0.7]{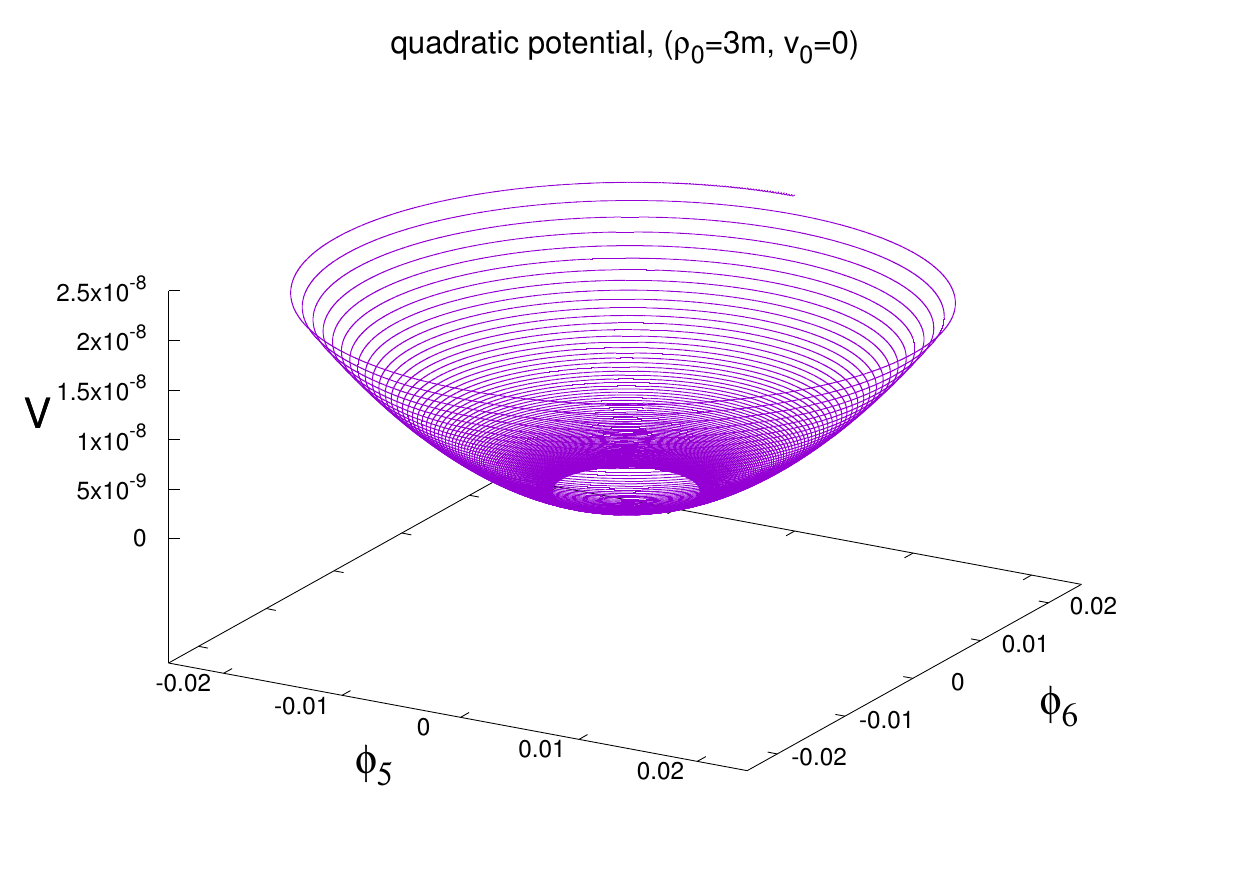}}
\caption{Numerical solution of the equations of motion (\ref{Einstein},\ref{eom3}) in a quadratic potential with initial $\r_0=3m, \vv_0=0$, over a period of two million Planck times.}
\label{qv0}
\end{figure}

\begin{figure}[h!]   
 \centerline{\includegraphics[scale=0.6]{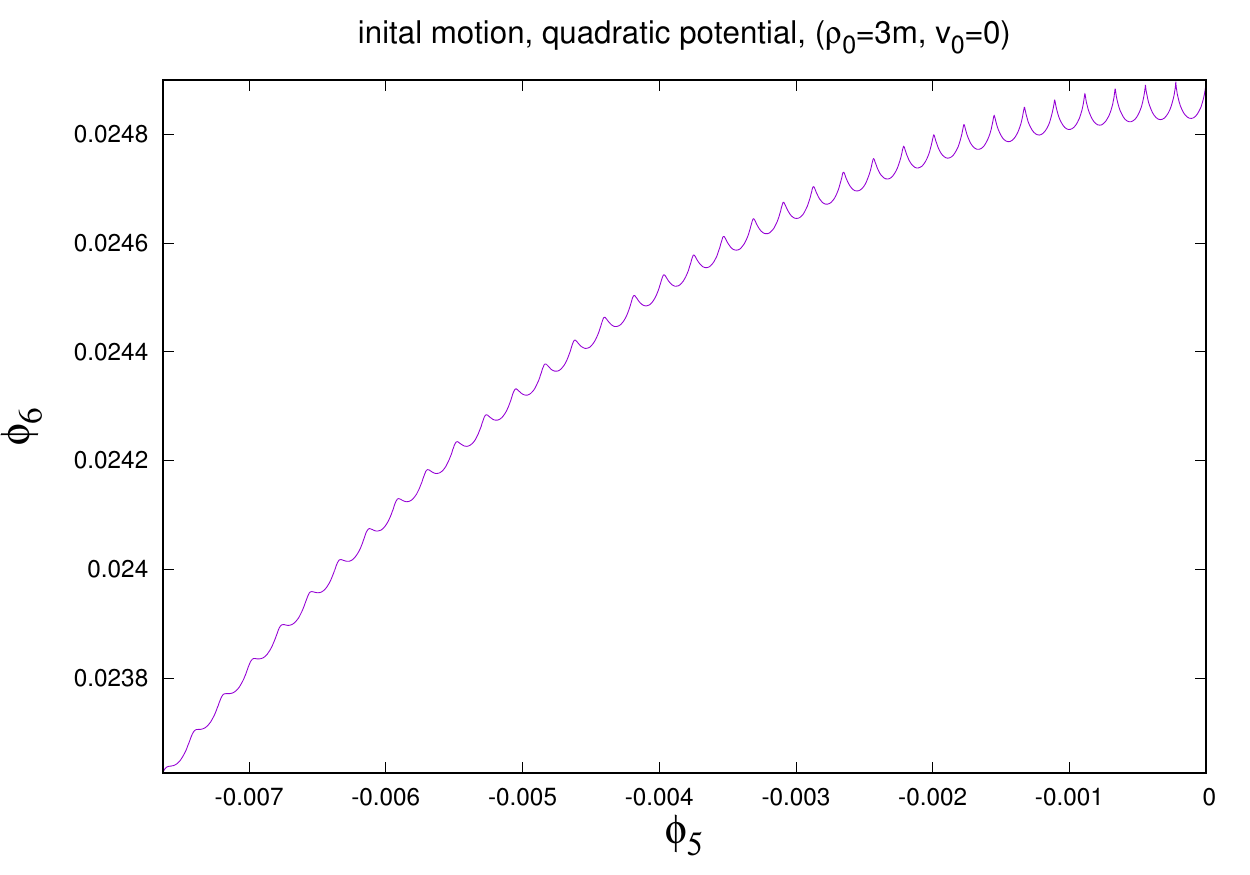}}
\caption{The initial trajectory of the previous figure, projected into the $\vph_5-\vph_6$ plane, which illustrates the initial
transient behavior that dies out after roughly two thousand Planck times.}
\label{qv0_start}
\end{figure}
   
   We begin with the quadratic potential \rf{quadratic}.  Inflationary behavior is generic, requiring only that $qB \gg m$ and
$\r_0 > m$.  For purposes of illustration we choose parameters to be motivated (for the Higgs potential) in the next section:
\beq
          m = 0.0083 ~~~,~~~ qB = 0.22
\eeq
in Planck units.   For the initial condition, we take
\beq
          \r_0 = 3 m  ~~~,~~~ \vv_0=0
\eeq
The trajectory in the space of $\vph_5,\vph_6,V(\r)$, obtained from these initial conditions by solving (\ref{Einstein},~\ref{eom3}) numerically
for Planck times $0\le t_P \le 2 \times 10^6$,  is displayed in Fig.\ \ref{qv0}.  There is some initial transient behavior which is shown 
in Fig.\ \ref{qv0_start}.  This behavior can be understood qualitatively:  From rest, the charged  ``particle'' tends to fall towards the minimum of the
potential, but as it falls there is a force orthogonal to the motion due to the external $B$-field.  The combination of forces induces the
transient oscillatory behavior seen in in Fig.\ \ref{qv0_start}, which dies out after $t_P=2000$ Planck times, i.e.\  the transient behavior is a very small fraction of the trajectory, and dies out rapidly.

\begin{figure*}[h!]   
 \centerline{\includegraphics[scale=1.2]{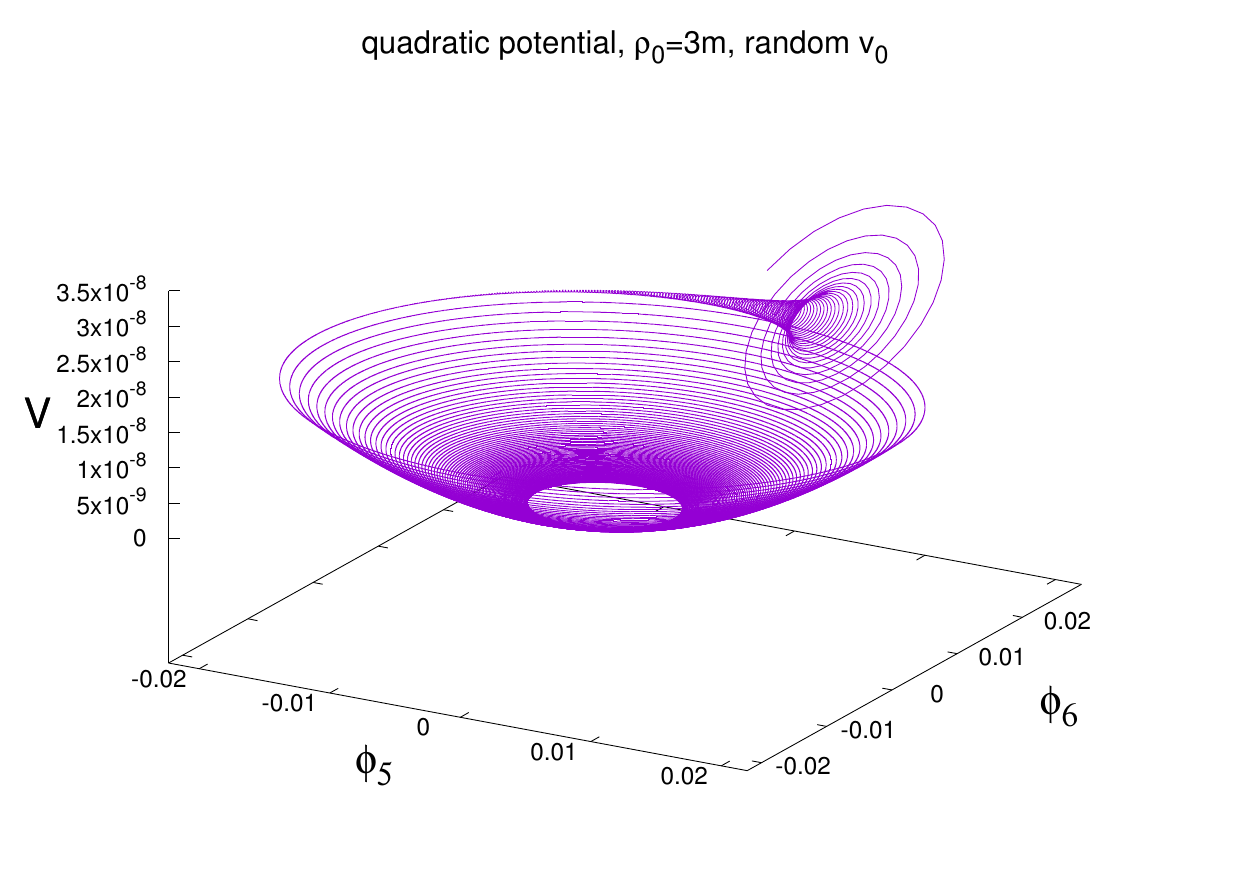}}
\caption{Same as Fig.\ \ref{qv0}, but with a random initial velocity $\vv_0$.  The transient behavior is clearly seen in this example.}
\label{qvran}
\end{figure*}

\begin{figure*}[h!]
\subfigure[~$H=\dot{a}/a$, quadratic potential]  
{   
 \label{quadhubble}
 \includegraphics[scale=0.5]{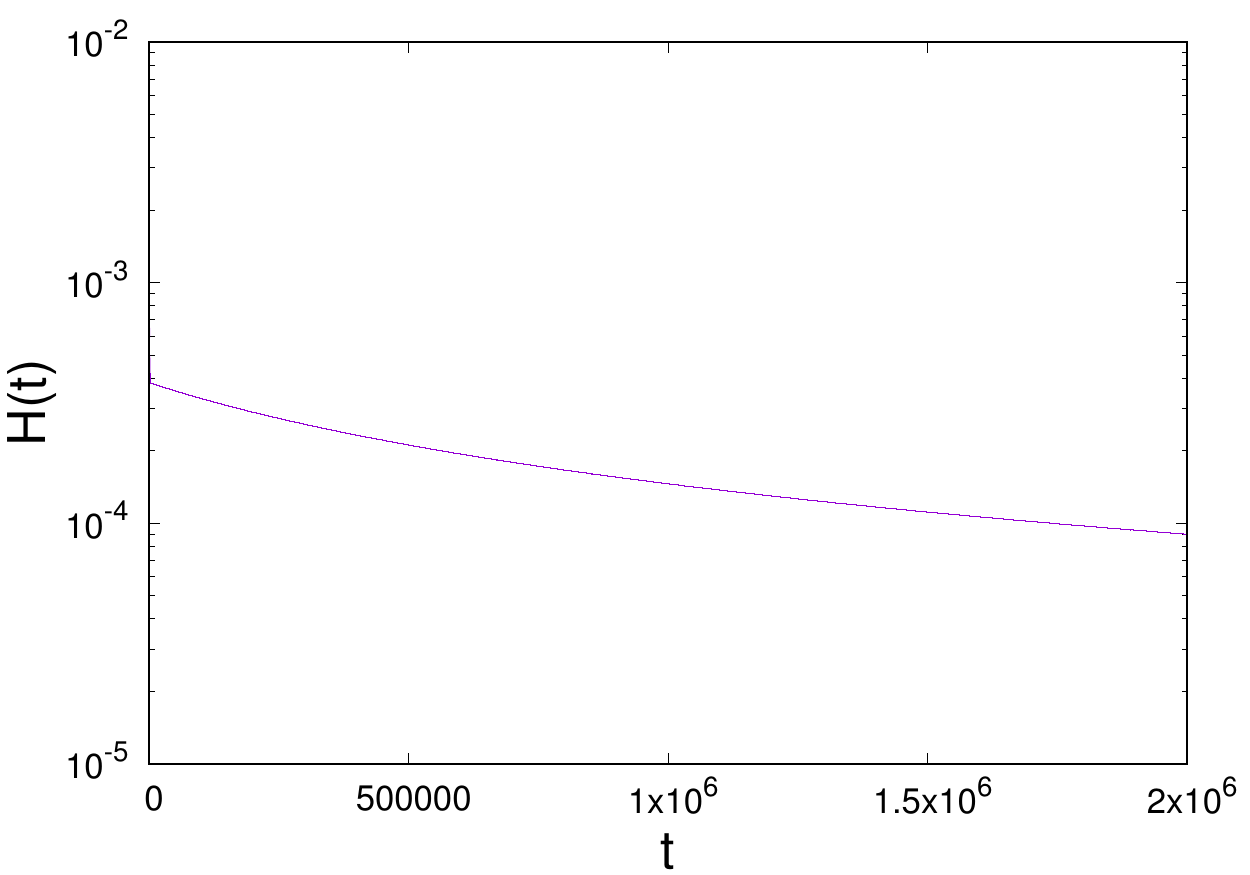}
}
\subfigure[~$\th(t)$, quadratic potential]  
{   
 \label{quadtheta}
 \includegraphics[scale=0.5]{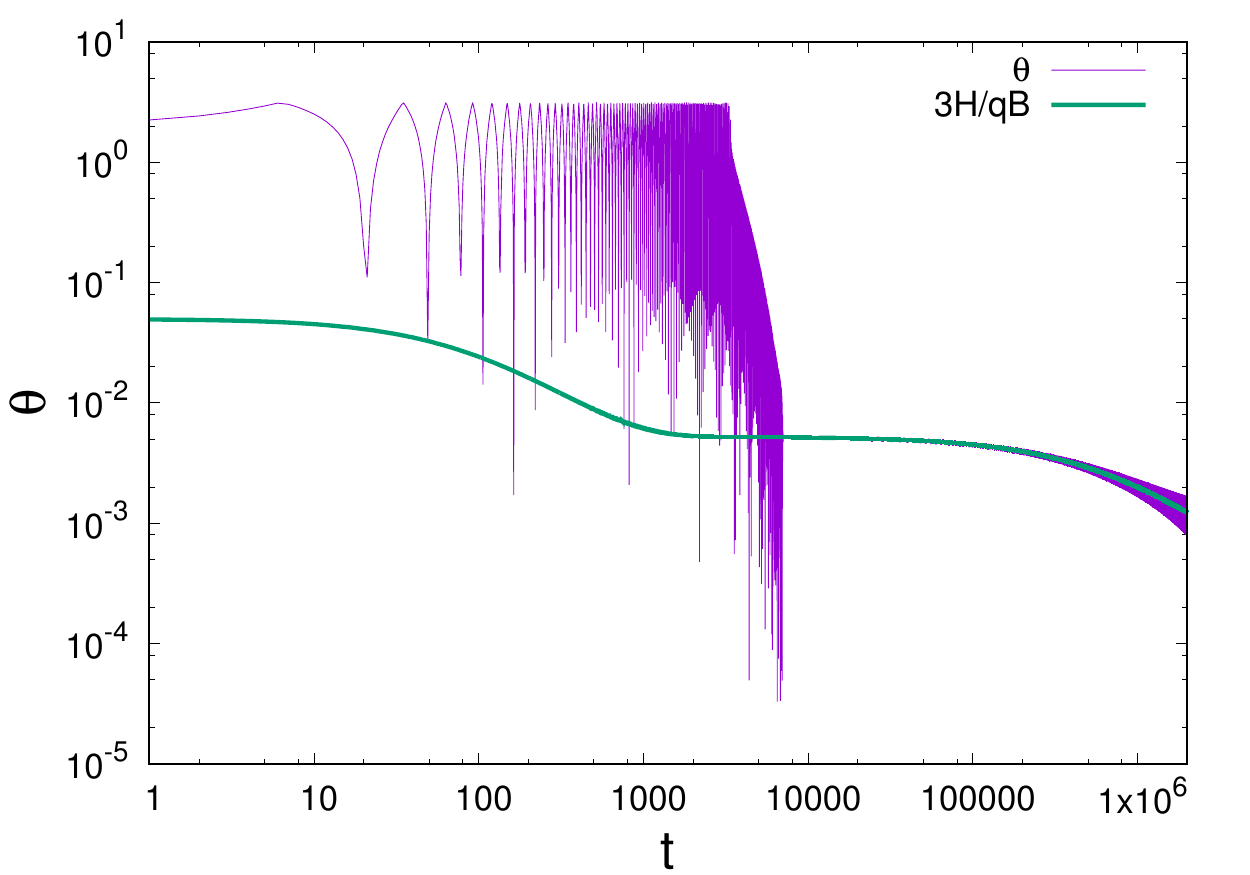}
}
\subfigure[~$\vv(t)$, quadratic potential]  
{ 
 \label{quadv}
 \includegraphics[scale=0.5]{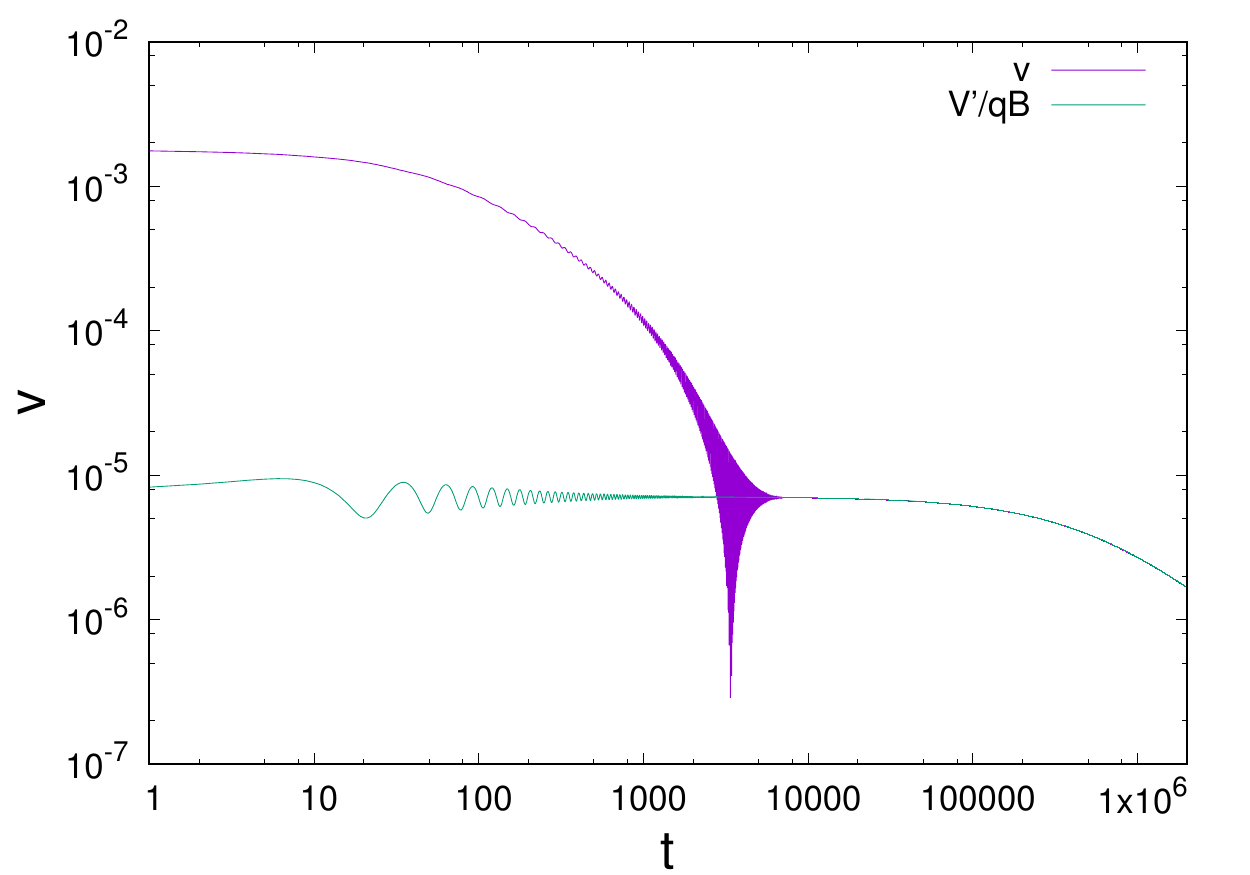}
}
\caption{For the example of the quadratic potential with a random initial $\vv_0$: (a) the expansion rate $H=\adot/a$. (b) Comparison of the computed $\th(t)$ in the $\vph_5-\vph_6$ plane with the slow spiral prediction \rf{slow0}. (c) Comparison of the speed $\vv(t)$ in the $\vph_5-\vph_6$ plane with the slow spiral prediction \rf{slow1}.}
\label{quad_evolution}
\end{figure*}

   If the universe starts out at non-zero $\rho_0$, then there is no particular reason that it also starts out at rest, so we may also consider
initial conditions with some random choice of $\vv_0$.  A solution of this kind, with $(\pa_t \vph_5,\pa_t \vph_6) = (8.4,15.6) \times 10^{-4}$ in Planck units, and all other parameters as above, is shown in Fig.\ \ref{qvran}, over a period of two million Planck times.  The transient behavior is much more noticeable in this case, but again it dies out after about ten thousand Planck times and the evolution settles into the slow spiral described in the previous section.   The behavior of $H=\adot/a$ is shown in Fig.\ \ref{quadhubble}, and we compare, in Figs.\ \ref{quadtheta} and \ref{quadv}, the behavior of
the computed $\theta(t), \vv(t)$ to the predictions of the slow spiral equations, \rf{slow0} and \rf{slow1} respectively.  The slow spiral equations clearly agree very well with the numerical solutions for $t_P>10,000$, i.e.\ after the transient behavior has died out.  The
number $N$ of e-foldings of course depends on the initial conditions and the assumed start of reheating.  For the trajectory shown
in these figures, $N=483$.

\begin{figure*}[h!]   
 \centerline{\includegraphics[scale=1.0]{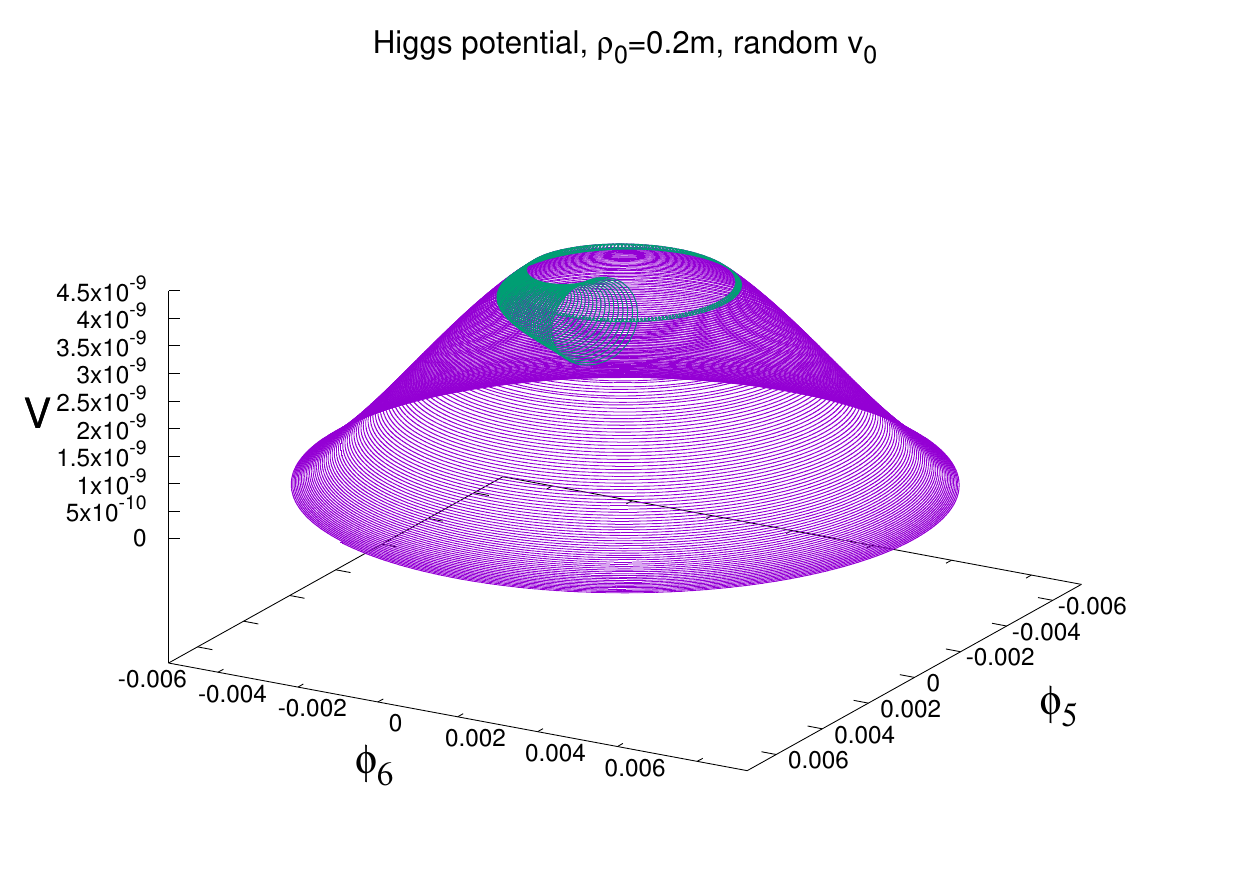}}
\caption{Numerical solution of the equations of motion (\ref{Einstein},\ref{eom3}) in a Higgs potential with initial $\rho_0=0.2 m$ and
random $\vv_0$, over a period of one million Planck times. The initial transient behavior is shown in green.}
\label{hran}
\end{figure*}

\begin{figure*}[h!]
\subfigure[~$H=\dot{a}/a$, Higgs potential]  
{   
 \label{higgshubble}
 \includegraphics[scale=0.5]{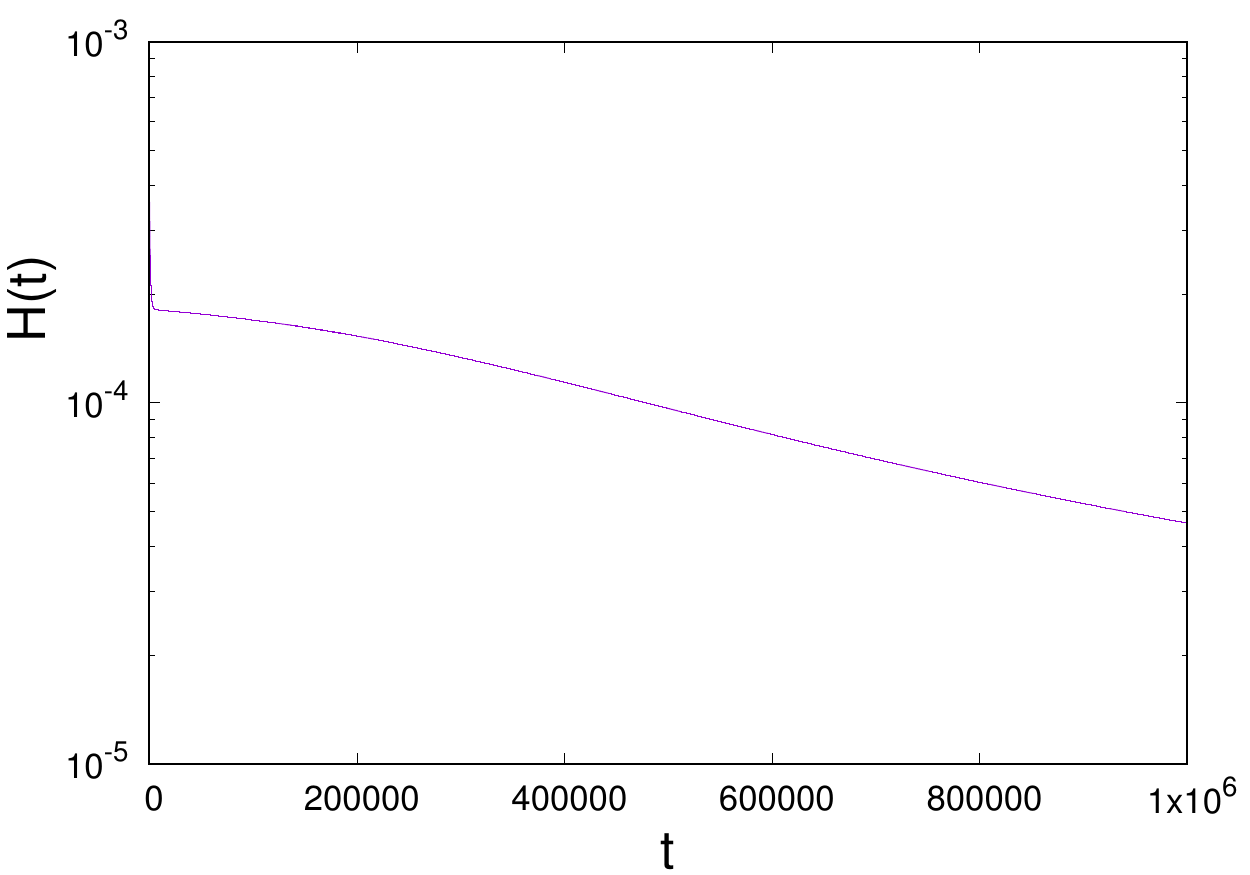}
}
\subfigure[~$\theta(t)$, Higgs potential]  
{   
 \label{higgstheta}
 \includegraphics[scale=0.5]{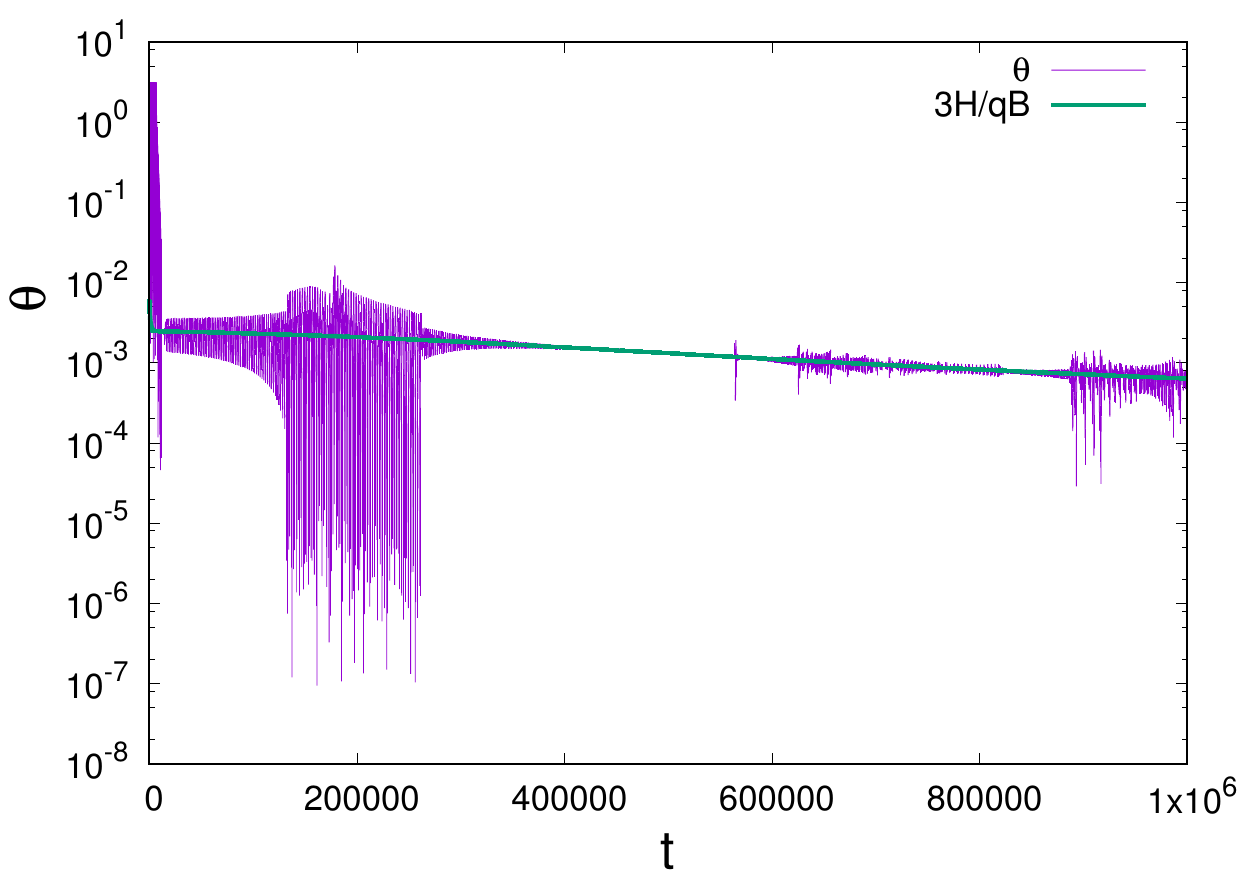}
}
\subfigure[~$\vv(t)$, Higgs potential]  
{ 
 \label{higgsv}
 \includegraphics[scale=0.5]{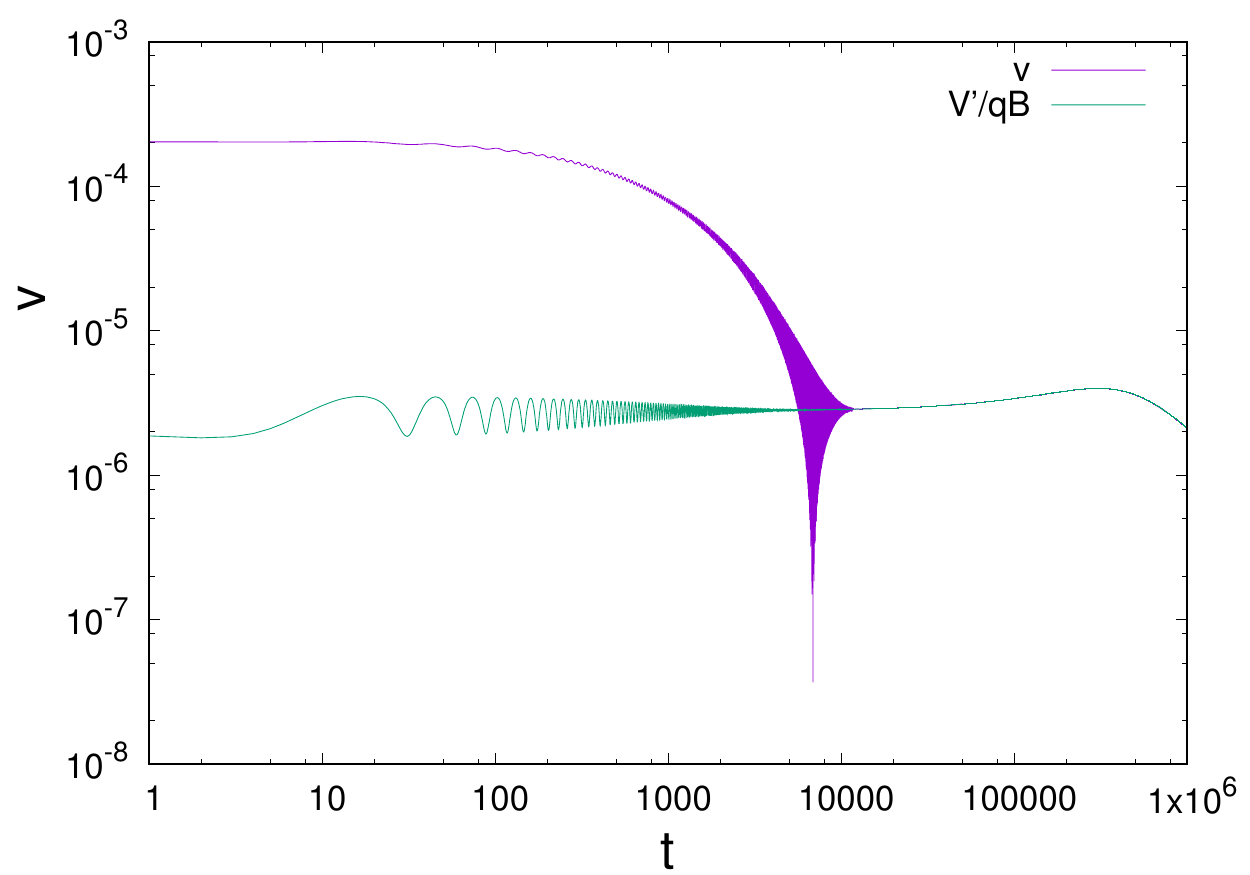}
}
\caption{For the example of the Higgs potential with a random initial $\vv_0$: (a) the expansion rate $H=\adot/a$. (b) Comparison of the computed $\th(t)$ in the $\vph_5-\vph_6$ plane with the slow spiral prediction \rf{slow0}. (c) Comparison of the speed $\vv(t)$ in the $\vph_5-\vph_6$ plane with the slow spiral prediction \rf{slow1}.}
\label{Higgs_evolution}
\end{figure*}

   Inflation is also generic for the Higgs potential when ${qB \gg m}$ and $m> \rho_0$, without much tuning beyond those conditions, for
any coupling $\l$.  As an illustration we carry out the computation for the Higgs potential with the same parameters 
$m = 0.0083, qB = 0.22$ as above, and an arbitrary choice of coupling $\l=1$.  The initial configuration is taken to be at $\r_0=0.2 m$,
with a random choice of initial velocity $(\pa_t \vph_5,\pa_t \vph_6) = (-1.37,1.50) \times 10^{-4}$ in Planck units.  The result, for a duration of one million Planck times, is shown in Fig.\ \ref{hran}, with the transient and later behaviors shown in different colors.  Again the transient behavior dies out beyond ten thousand Planck times.  The Hubble variable $H=\adot/a$ vs.\ time, is shown in Fig.\ \ref{higgshubble}, and the
computed $\theta(t),\vv(t)$, compared to the slow spiral predictions  \rf{slow0} and \rf{slow1}, are shown in Figs. \ref{higgstheta} and
\ref{higgsv} respectively.  Once again it is clear that the slow spiral is an attractor, at least under the stated conditions $qB \gg m$ and 
$m> \rho_0$.  For the evolution shown, where the trajectory is computed up to $\r_{final}={7\over 8} m$, the number of e-foldings is
$N=104$.

\section{Fluctuation spectrum}

   We now consider the equations of motion to first order in fluctuations around the homogeneous isotropic background.  Denote
\beq
\vph^s(x,t) = \vph^s_0(t) + \d\vph^s(x,t) \ ,
\eeq
where $\vph^s_0(t), a(t)$ are solutions of the spatially homogeneous equations (\ref{Einstein},~\ref{eom3}).   In a standard
slow-roll scenario with a single inflaton field, it is possible to choose a gauge (essentially a choice of time coordinate) such that
$\d \vph = 0$.  The transformed inflaton field is
\beq
       \vph'(x',t') = \vph(x,t) 
\eeq
where $x'=x$ and
\beq
t'(x,t) = \vph_0^{-1}[\vph(x,t)]
\eeq
Since $t'(x,t)$ is, by definition, the time at which $\vph_0(t')=\vph(x,t)$, it is clear that $\vph'(x',t') = \vph_0(t')$ and therefore,
by construction, $\d \vph(x,t) = 0$ in this (co-moving) gauge.

    With two inflaton components, it is not possible to transform to $\d \vp = 0$ exactly.  However, in the slow-spiral regime it is 
possible to come close to this, under certain conditions.  The reason one cannot find a time coordinate such that $\vp'(x',t')=\vp_0(t')$
is simply because $\vp_0(t)$ defines a line-like object, i.e.\ a spiral, while $\vp(x,t)$ lies in a plane.  But one can still choose
$t'(x,t)$ to be such that the modulus of
 \beq
  \d \vp(x,t') \equiv \vp(x,t)-\vp_0(t')
\eeq 
is minimized.  In this ``minimal'' gauge $|\d \vp(x',t')|$ is the distance in the $\vp$-plane  between $(\vph^5(x,t),\vph^6(x,t))$ and the nearest point on the spiral.   This distance can be no greater than half the distance between neighboring arcs of the spiral.  From eqs.\ \rf{slow2} and \rf{slow3}, this upper bound on the minimal distance is
\bea
           \Delta \rho &=& \oh  \left|{d\rho \over d\a}\right| 2\pi  \non \\
                       &=& {3\pi H \over qB} \rho
\eea
 So in minimal gauge the magnitude of $\d \vph$ is as small as it can be, with a fractional deviation $\d \r/\r$ of order $H/qB$.  If 
 the Hubble expansion rate $H$ is very much smaller than $qB$, then in this gauge the fluctuations away from 
$\vp_0(t)$ can be neglected by comparison to the $\d \phi^A$, which give rise to fluctuations in the metric.  We will assume that this $H \ll qB$ condition is satisfied, and that in an expansion of the action $S = S_0[\vp_0,\phi^A_0] + \d S$ to second order in the fluctuations the terms involving $\d \vp$ can be ignored relative to terms quadratic in $\d \phi^A$.

    At this point we can simply follow the approach of Maldacena \cite{Maldacena:2002vr}, 
who derives the power spectrum in a comoving gauge where the fluctuations of a single inflaton field vanish.  The only modification is to replace the single inflaton field expression $(\pa_t \vph_0)^2$ by $\vv^2=\pdot_0^s \pdot_0^s$.  Of course, in quantizing perturbations around the Friedmann-Lemaitre metric one must keep in mind that $g_{\m\n}$ is the induced metric \rf{induce}.  However, in a minimal gauge with $\d \vp \approx 0$, the fluctuation $\d S$ in the action depends on the fluctuations of $\phi^A=\phi^A_0 + \d \phi^A$ only through the induced metric, and therefore we have
\bea
     Z &=& \int D \d \p^C \exp\Bigl[-(S_0 + \d S[\pa_\m \p^A \eta_{AB} \pa_\n \p^B])\Bigr]  \non \\
        &=& \int D \d \p^C \int Dg_{\a\b} \d[g_{\m\n} - \pa_\m \p^A \eta_{AB} \pa_\n \p^B] e^{-(S_0+\d S[g_{\m\n}])}  \non \\
        &=&  \int Dg_{\a\b}  M[g] e^{-(S_0 + \d S[g_{\m\n}])} \ ,
\eea
where
\beq
         M[g] =  \int D \d \p^C \d[g_{\m\n} - \pa_\m \p^A \eta_{AB} \pa_\n \p^B] \ .
\eeq
The measure factor $M[g]$ will not contribute at the quadratic order in metric fluctuations considered here.  Then, following
\cite{Maldacena:2002vr}, we obtain the familiar result for the dimensionless power spectrum of the scalar curvature fluctuations
\beq
           P_R(k) = {1\over 4\pi^2} {H^4 \over \vv^2} \ .
\eeq
For the spectral index $n_s$, with $N(t) = \int dt H$ the number of e-foldings, we have \cite{Baumann:2009ds}
\bea
     n_s-1  &=& {d \ln P_R \over d \ln k} \non \\
      &\approx&  \left( 2 {d\ln H^2 \over dN} - {d \ln \vv^2 \over dN}\right)\left(1 -\oh {d\ln H^2 \over dN}\right) \non \\
      &=& {1\over H}\left( 2 {d\ln H^2 \over dt} - {d \ln \vv^2 \over dt}\right)\left(1 -{1\over 2H} {d\ln H^2 \over dt}\right) \ , \non \\
\eea
and for the $r$-parameter
\bea
          r &=& {P_{grav} \over P_R} = {{2\over \pi^2}{H^2 \over M_P^2} \over {1\over 4\pi^2}{H^4 \over \vv^2}} \non \\
            &=& 8 {\vv^2 \over M_P^2 H^2} \ ,
\eea
where $M_P$ is the Planck mass.   
Now we apply the slow spiral equations \rf{slow1} and \rf{slow2}, using also the approximation to the Friedmann equation 
$H^2 \approx  V(\r)/3M_P$.  Then we have
\bea
          P_R^{1/2} &=& {1 \over 6\pi} {qB \over M_P^2} {V(\r) \over |V'(\r)|} \non \\
          n_s -1 &=& -{6\over (qB)^2}\left( {V'^2 \over V} - V''\right)\left(1 + {3\over 2}{1\over (qB)^2} {V'^2 \over V}\right) \non \\
          r &=& {24 \over (qB)^2} {V'^2 \over V}  \ ,
\eea
and from the Planck 2015 \cite{Ade:2015lrj} and earlier data it is known that
\bea
          P_R^{1/2} &\approx& 4.6 \times 10^{-5} \non \\
          n_s - 1  &\approx& -0.04 \non \\         
          r &<& 0.1
\label{data}
\eea
at the pivot scale $k=0.05$ Mpc${}^{-1}$.
          
\subsubsection{Quadratic Potential}

   For the quadratic potential $V(\r) = \oh m^2 \r^2$ these expressions become
\bea
     P_R^{1/2} &=& {1 \over 12 \pi} {qB \over M_P^2} \r \non \\
     n_s - 1 &=& -{6 m^2 \over (qB)^2} \non \\
     r &=& 48 \left(m \over qB\right)^2 .
\eea
But from the Planck data on the spectral index, $n_s-1 \approx -0.04$, these equations imply that
$r \approx 0.32$, which is incompatible with the Planck data for $r$.  On these grounds, the quadratic potential is
ruled out.

\subsubsection{Higgs Potential}

   For the Higgs potential $V(\r)=\l (\r^2 - m^2)^2$ the corresponding expressions are
\bea
     P_R^{1/2} &=& {1 \over 24\pi} {qB \over M_P^2} {|\r^2 -m^2| \over \r} \non \\
         n_s - 1  &\approx& -{24 \over (qB)^2} \l(\r^2 + m^2) \non \\
           r &=& 384 { \l \r^2 \over (qB)^2} \ .
\label{higgs_eqs}
\eea
The coupling $\l$ can be entirely eliminated from these equations by writing
\beq
       m = m_0 \l^{-1/4} ~~,~~   qB = (qB)_0  \l^{1/4}  ~~,~~ \r = \a m \ ,
\eeq
and then the constants $m_0,(qB)_0$, and $\a$ at the pivot scale, are determined, given $P_R, n_s$ and $r$.  Unfortunately we only have an upper bound for $r$, so the best one can do so far is to determine $\m,b,\a$ as a function of $r<0.1$.  For sufficiently small $r$ we find the approximate solution
\bea
           m  &\approx& \sqrt{24\pi} \left({P_R \over 384}\right)^{1/4} \left({r \over \l}\right)^{1/4} M_P \non \\
           \r_* &\approx& \sqrt{24 r \over 384 (1-n_s)} m \non \\
                  &=& 24 \sqrt{\pi \over 384(1-n_s)}  \left({P_R \over 384}\right)^{1/4} {r^{3/4} \over \l^{1/4}} M_P \non \\
          qB &\approx&  24 \sqrt{\pi \over (1-n_s)} \left({P_R \over 384}\right)^{1/4} r^{1/4} \l^{1/4} M_P \ ,
\label{powers}
\eea
where $\rho_*$ is $\rho$ at the pivot scale.  To get some numerical feeling for this, suppose $r$ is at the experimental upper limit
$r=0.1$ and $\l=1$.  Then, using the data \rf{data}, we find from \rf{higgs_eqs}
\bea
   \r_* &=&  3.6 \times 10^{-3}  ~ M_P  \non \\
   m &=& 8.3 \times 10^{-3} ~ M_P \non \\
    qB &=& 0.22  ~ M_P \ .
\label{lambda1}
\eea
We can also compute the running of the spectral index
\bea
        {dn_s \over d\ln k} &=& {d n_s \over dN}{dN \over d\ln k} = {1\over H}  {d n_s \over dt}(1+\e)  \non \\
             &\approx&  - {1\over H} {96 \l \over (qB)^2} \r {d\r \over dt} \non \\
             &=& 288 \l^2 {\r^2 (\r^2-m^2) \over (qB)^4}
\eea
which, evaluated at $\r=\r_*$ in our $r=0.1$ example, is
\beq
 {dn_s \over d\ln k} = -0.9 \times 10^{-4} \ .
\eeq
Note that, from \rf{powers}, this value is $\l$-independent, and can be compared to the Planck 2015 \cite{Ade:2015lrj} estimate of
 \beq
 {dn_s \over d\ln k} = -(3.3 \pm 7.4) \times 10^{-3} \ .
\eeq
  
\section{e-foldings in the Higgs slow spiral}

    We have seen that after a brief transient, the solution of the homogenous equations of motion (\ref{Einstein},\ref{eom3}) settles into a trajectory which is well described by the slow spiral equations of motion derived in section \ref{spiral}.  We have also seen, in the previous section, that the constants $m$ and $qB$ are determined, for the Higgs potential, from the CMB data for $P_R, n_s, r$, and the Higgs coupling $\l$.   Moreover, $m$ and $qB$ depend on simple fractional powers of $\l,r$, shown in \rf{powers}.  From this power dependence, and the slow spiral equations of motion, it is quite easy to see that the number $N$ of e-foldings is independent
of $\l$ and $r$.   For the Higgs potential, with the minimum at $\r=m$, this number is determined by the known values of $P_R, n_s$, and the (as yet unknown) fractions of $\r/m$ at the beginning and end of inflation.  

\begin{figure}[h!]   
 \centerline{\includegraphics[scale=0.7]{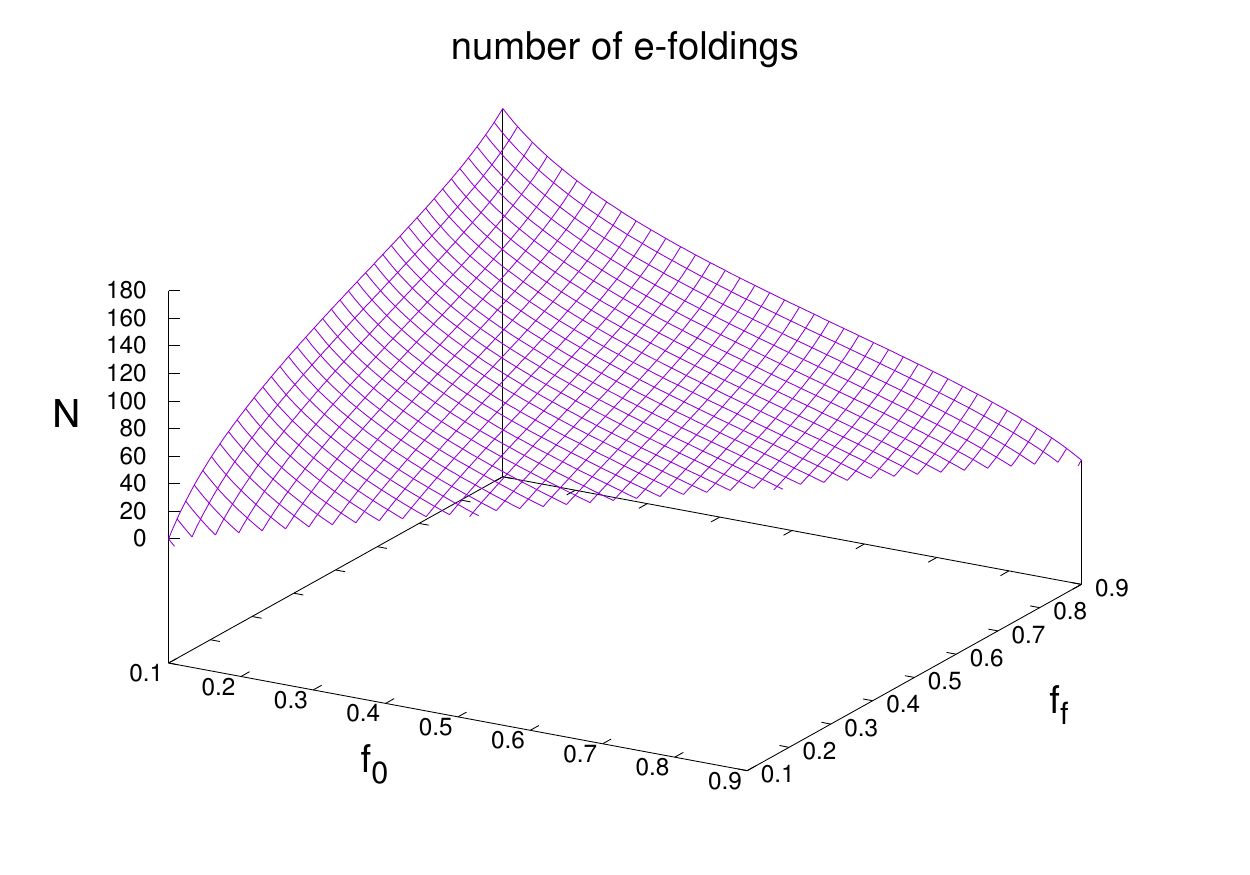}}
\caption{The number of e-foldings $N$, in the Higgs potential case, as a function of the inflaton field $\rho=|\vp|$ at the start and end of inflation.  The latter are displayed as fractions $f_0=\rho_{initial}/m$ and $f_f=\rho_{final}/m$ of the minimum of the Higgs potential at 
$\r=m$.}
\label{efold}
\end{figure}

      Let $m_0, (qB)_0$ denote the values of $m,qB$ shown in \rf{lambda1}, which were determined from the Planck data assuming 
$\l=1,r=0.1$.  Then from \rf{powers}
\bea
           m &=& m_0 \l^{-1/4} \left({r\over 0.1}\right)^{1/4}  \non \\
           qB &=& (qB)_0  \l^{1/4} \left({r\over 0.1}\right)^{1/4}
\eea
and also denote
\beq
          \r = R \l^{-1/4} \left({r\over 0.1}\right)^{1/4}
\eeq
Substituting these expressions into the slow-spiral equations of motion, we find
\bea
            H^2 &=& {1\over 3M_P} (R^2-m_0^2)^2   \left({r\over 0.1}\right) \non \\
                    &=& H_0^2 \left({r\over 0.1}\right) \non \\
            \vv &=& {4 R(R^2-m_0^2) \over (qB)_0} \left({r\over 0.1}\right)^{1/2}\non \\
                  &=& \vv_0 \left({r\over 0.1}\right)^{1/2} \non \\
           {dR \over dt} &=& -3H_0 {4 R(R^2-m_0^2) \over (qB)_0^2}  \left({r\over 0.1}\right)
\eea
Then the number of e-foldings from the beginning of inflation at $t=0,R=R_0$ to the end of inflation at $t=t_f, R=R_f$ is
\bea
       N &=& \int_0^{t_f} dt H = \int_{R_0}^{R_f} dR \left(dR \over dt\right)^{-1} H \non \\
          &=& {(qB)_0^2 \over 12} \int_{R_0}^{R_f} {dR \over R(m_0^2-R^2)} \non \\
          &=&  {(qB)_0^2 \over 24 m_0^2} \log\left[ {R_f^2 (m_0^2-R_0^2) \over R_0^2 (m_0^2-R_f^2)}\right]        
\eea
Note that both $\l$ and $r$ have dropped out of this expression.  Since $(qB)_0$ and $m_0$ are determined from the Planck data, the
number of e-foldings depends entirely on the fractions
\bea
          f_0 &=&  {\r_{initial} \over m} = {R_0 \over m_0}  \non \\
          f_f &=&   {\r_{final} \over m} = {R_f \over m_0} 
\eea
in terms of which
\bea
    N &=&  {(qB)_0^2 \over 24 m_0^2} \log\left[ {f_f^2 (1-f_0^2) \over f_0^2 (1-f_f^2)}\right]  \non \\
       &=& 29.3  \log\left[ {f_f^2 (1-f_0^2) \over f_0^2 (1-f_f^2)}\right] 
\eea
Although the number of e-foldings is  $\l,r$-independent, the period of inflation does have an $r$-dependence, increasing as $r$
decreases like $1/\sqrt{r}$:
\bea
           t_f &=&  \int_{R_0}^{R_f} dR  \left(dR \over dt\right)^{-1} \non \\
                &=& {(qB)_0^2 \over 4 \sqrt{3}} M_P \left({r\over 0.1}\right)^{-1/2} \int_{R_0}^{R_f} {dR \over R(m_0^2-R^2)} \non \\
                &=& {(qB)_0^2 \over 8 \sqrt{3}} {M_P \over m_0^4}\left({r\over 0.1}\right)^{-1/2}
                \bigg\{ \log\left[ {R_f^2 (m_0^2-R_0^2) \over R_0^2 (m_0^2-R_f^2)}\right] m \non \\
                & & \qquad  - {m_0^2(R_f^2-R_0^2) \over (m_0^2-R_0^2)(m_0^2-R_f^2)} \bigg\}
\eea                
Finally, although we do not know the initial and ending points $\r_{initial},\r_{final}$, it is clear that $\r_{initial}$ must be less
than $\r$ at the pivot momentum.  From \rf{powers}
\beq
           \r_* = R_*  \l^{-1/4} \left({r\over 0.1}\right)^{3/4}
\eeq
where $R_*=3.6\times 10^{-3} M_P$ is the value of $\r_*$ shown in \rf{lambda1} for $\l=1,r=0.1$.  Then defining, for $r \le 0.1$,
\bea
           f_* &=& {\r_* \over m} = {R_* \over m_0} \left({r\over 0.1}\right)^{1/2} \non \\
                &=&  0.43 \left({r\over 0.1}\right)^{1/2}
\eea
it is necessary that $f_0 < f_*$.   The number of e-foldings $N$ vs.\ the fractions $f_0,f_f$ are displayed in Fig.\ \ref{efold}

\section{Excitations}

   It is worth mentioning, for the sake of completeness, some intriguing effects of the external four-form gauge field if the field strength persists into the late universe.  It is found, when the
inflaton field is quantized in the presence of the constant background four-form field strength, that there is an analogy to
ordinary Landau levels, and the spectrum of the quantized field is
\bea
E &=& \sum_k \left\{ \sqrt{k^2 +\oq q^2 B^2 + m^2} \bigg(n_1(k) + n_2(k)\bigg) \right. \non \\ 
   & & \left. + \oh qB\bigg(n_1(k) - n_2(k)\bigg) \right\} + E_0 \; ,
\label{spectrum}
\eea
where $n_1(k),n_2(k)$ are occupation numbers, $E_0$ is the ground state energy, and the sum runs over
momenta with non-zero occupation numbers.  Excitations with definite energy and momentum satisfy
one of two dispersion relations
\bea
      E_1(k) &=& \sqrt{k^2 + M^2 + m^2} + M\; , ~~~\mbox{and} \non \\
      E_2(k) &=& \sqrt{k^2 + M^2 + m^2} - M \; ,
\eea
where $M=\oh qB$.   It can be shown that these two types heavy/light excitations propagate like ordinary massive 
particles with inertial mass $M'=\sqrt{M^2+m^2}$, i.e.\ with group velocity
$\vv = k/\sqrt{k^2 + M^2 + m^2}$,
but interact gravitationally with gravitational masses
$M' \pm M$.  Thus there is an apparent violation of the Principle of Equivalence, due to the interaction with the 
external field strength of the four-form gauge field.   For details, cf.\  \cite{Greensite:2016dhu}.

\bigskip

\section{Many braneworlds}

    Having introduced a braneworld scenario in which the braneworld can interact with an external four-form abelian gauge field, one can of course formulate the obvious generalization:  a higher-dimensional universe containing many braneworlds, interacting with one another via the four-form gauge field, which itself has dynamics and wave propagation in the bulk.  The action would be
\bea
    \lefteqn{S =} 
     & & \int \prod_a d\p^a F_{bcdef}[\p]  F^{bcdef}[\p]  \non \\
    & & + \sum_n \left\{ {q_0\over 4!} \int d\p^a \wedge d\p^b \wedge d\p^c \wedge d\p^d ~ A_{abcd}[\p(x_n)] \right. \non \\
    & & +  S_{infl}[\{\p^s(x_n)\},\sqrt{g(x_n)}]+S_{EH}[g_{\m\n}(x_n)]  \non \\
    & & \left. + S_{SM}[\{\Phi(x_n)\},\sqrt{g(x_n)}]  \right\} \non \\
\eea
where the sum is over braneworlds, $\phi^a(x_n)$ are the coordinates of the $n$-th braneworld in the bulk, $S_{infl},~S_{EH},~S_{SM}$ are the inflaton,
Einstein-Hilbert, and standard model actions, with $\{\Phi(x_n)\}$ the set of standard model fields living on the
${n\mbox{-th}}$ braneworld,
and $g_{\m\n}(x_n)=\pa_\m \p^A(x_n) \eta_{AB} \pa_\n \p^B(x_n)$ the induced metric of the {$n$-th} braneworld.  At the quantum level
one could even speculate, going well beyond the conjecture raised in this article, that our $D=4$ dimensional universe was created in a strong external gauge field via pair production of a three brane-antibrane pair.  

   We are content to make these speculations, but will not pursue them here.  A consistent formulation would probably require making some connection to string theory.  
  
\section{Conclusions}  
  
    We have pointed out that if the universe is to be thought of as a three-brane propagating in a higher dimensional space, then
it is natural for that three-brane to couple to a four-form gauge field in the bulk.  If the very early universe were exposed to a strong external field strength due to the four-form gauge field, then it seems that the resulting ``slow spiral'' of the inflaton field would solve one of the main problems associated with small field inflation in a Higgs (or other hilltop) potential, namely, the need for fine-tuning the initial value of the inflaton field and its time derivative.   We reserve questions regarding reheating and non-Gaussianity in this scenario for later investigation.
  
\bigskip
  
\acknowledgments{This research is supported by the U.S.\ Department of Energy under Grant No.\ DE-SC0013682. }

\bibliography{bff}

\end{document}